\documentclass[amssymb,superscriptaddress]{revtex4}
\usepackage{amsmath,amsfonts,amssymb,amstext}
\usepackage{bbm}
\usepackage{color}

\font\sqi=cmssq8
\newcommand{\ba}{\begin{array}}
\newcommand{\ea}{\end{array}}
\newcommand{\be}{\begin{equation}}
\newcommand{\ee}{\end{equation}}
\newcommand{\beqs}{\begin{eqnarray}}
\newcommand{\eeqs}{\end{eqnarray}}
\newcommand{\bea}{\begin{eqnarray}}
\newcommand{\eea}{\end{eqnarray}}

\def\DR{\rm I\kern-1.45pt\rm R}
\def\DC{\kern2pt {\hbox{\sqi I}}\kern-4.2pt\rm C}
\def\iprime{\mathcal{I}}
\def\ical{\mathcal{I}}
\def\sfrac#1#2{{\textstyle\frac{#1}{#2}}}
\def\und{\qquad\textrm{and}\qquad}
\def\Z{\mathbb Z}
\def\R{\mathbb R}

\begin{document}

\begin{flushright}
ITP--UH--14/10
\end{flushright}

\title{Invariants of the spherical sector in conformal mechanics}

\author{Tigran Hakobyan}
\email{hakob@ysu.am} \affiliation{Yerevan State University, 1 Alex
Manoogian, 0025 Yerevan, Armenia} \affiliation{Yerevan Physics
Institute, 2 Alikhanyan Br., 0036 Yerevan, Armenia}
\author{Olaf
Lechtenfeld} \email{lechtenf@itp.uni-hannover.de}

\affiliation{ Leibniz Universit\"at Hannover, Institut f\"ur
Theoretische Physik, Appelstr. 2, D-30167 Hannover Germany}
\author{Armen Nersessian}
\email{arnerses@ysu.am} \affiliation{Yerevan State University, 1
Alex Manoogian, 0025 Yerevan, Armenia}
\author{Armen Saghatelian}
\email{asaghat@physic.ysu.am} \affiliation{Yerevan State University,
1 Alex Manoogian, 0025 Yerevan, Armenia}

\begin{abstract}
A direct relation is established between the constants of motion for
conformal mechanics and those for its spherical part. In this way we find
the complete set of functionally independent constants of motion for the
so-called cuboctahedric Higgs oscillator, which is just the spherical part
of the rational $A_3$ Calogero model (describing four Calogero particles
after decoupling their center of mass).
\end{abstract}
\maketitle

\section{Introduction}
Recently, there has been new interest in so-called ``conformal mechanics''.
This term denotes a system whose Hamiltonian ${H}$,
together with the dilatation generator ${D}$ and the
generator ${K}$ of conformal boosts forms, with respect to
Poisson brackets, the conformal algebra $so(1,2)$:
\be \label{so120}
\{ {H} , {D}\}= 2 {H}, \qquad
\{ {K} , {D}\}=-2 {K}, \qquad
\{ {H} , {K}\}=   {D}. \ee
Such system can always be presented in the form \cite{hkln}
\be
\label{so2}
{ D}=p_r r ,\qquad
{ K}=\frac{r^2}{2},\qquad
{ H}=\frac{{p}^2_r}{2}+ \frac{\iprime(u)}{2r^2},
\ee
where the radial coordinates $(r,p_r)$ and the angular coordinates $(u^\alpha)$
obey the basic Poisson brackets
\be
\{p_r, r\}=1,\qquad
\{u^{\alpha}, p_r\}=\{u^{\alpha}, r\}=0, \qquad
\{u^\alpha, u^\beta\}=(\omega^{-1})^{\alpha\beta}(u).
\ee
The spherical (or angular) part of the Hamiltonian ${H}$,
\be \label{casimir}
\iprime=4{KH}-{D}^2,
\ee
is the Casimir element of \eqref{so120}
and, hence, commutes with all generators but also defines a constant of motion
of the Hamiltonian ${H}$.

The spherical part of the conformal mechanics, determined by
\be
\omega_0:=\sfrac12 \omega_{\alpha\beta} du^\alpha{\wedge}du^{\beta}
\und \iprime,
\ee
may be considered as a Hamiltonian system by itself.
We refer to it as "spherical mechanics" throughout the paper.
It is obvious that integrability of the initial conformal mechanics
leads to integrability of the ``spherical mechanics'' $(\omega_0,
\iprime)$, and vice versa. It is also evident that the constants of
motion of the spherical mechanics are constants of motion for the
conformal mechanics. Yet, the inverse is generally not true, although
there should be a way to construct the ``spherical'' constants of motion
out of the "conformal" ones. This is the problem we address in this paper.

In \cite{hkln} some of us began a study of spherical mechanics.
It is relevant for investigations of the Calogero model \cite{calogero69,moser}
and its various extensions and generalizations \cite{algebra}
(for a recent review see \cite{polychronakos}).
Furthermore, the spherical mechanics  of the rational $A_N$ Calogero model
defines the multi-center (Higgs) oscillator system on the $N{-}1$-sphere
\cite{cuboct}. The well-known series of hidden constants of motion found
by Wojcechowski \cite{woj83} for the Calogero model has a transparent
explanation in terms of spherical mechanics, and its analog exists in any
integrable conformal mechanical system \cite{hkln}.
Even in the simplest case of $N{=}2$, the one-dimensional spherical mechanics
of the $A_2$ Calogero model shed light on a global aspect of Calogero models,
by elucidating the non-equivalence of different quantizations of the Calogero
model \cite{feher}.
The ${\cal N}{=}4$ superconformal generalizations of the rational $A_2$
Calogero model, constructed via supersymmetrization of spherical mechanics
\cite{bks}, yielded a scheme for lifting any ${\cal N}{=}4$
supersymmetric mechanics to a $D(1,2|\alpha)$ superconformal one \cite{hkln}.
Finally, a formulation in terms of action-angle variables \cite{lny} led
to the equivalence of the rational $A_2$ and $G_2$ Calogero models and provided
restrictions on the ``decoupling'' transformation which maps the Calogero model
to the free-particle system considered in \cite{decoupling,anton}.

Directly relevant for the task of the present paper,
it was in fact demonstrated in \cite{hkln} that all information
on a conformal mechanics system is encoded in its spherical part.
In particular, the ``conformal'' constants of motion with even conformal
dimension were shown to induce constants of motion for $(\omega_0, \iprime)$.
However, the authors were unable to find the ``spherical'' constants of motion
induced by the odd-dimensional initial constants of motion.
In the following, we are going to solve this problem with the help of $so(3)$
representation theory.

The paper is arranged as follows:
In Section~II, following but extending \cite{hkln},
we relate the symmetries of conformal mechanics to the particular system
of differential equations on the spherical phase space.
The analysis is simplified by the use of $so(3)$ representations,
which clarifies the origin of the spin operators appearing in the final system.
In the Section~III we construct a series of the constants of
motion for the spherical mechanics, which is induced by the constants of
motion (of any conformal dimension) for the conformal system.
In Section~IV we apply our method to the rational $A_3$ Calogero model
and derive the complete set of functionally independent constants of motion
for the cuboctahedric Higgs oscillator.

\section{The spherical part of conformal mechanics ("spherical mechanics")}
In this section, we  relate the constants of motion of the conformal mechanics (\ref{so2}) with certain differential equations on the phase space of the associated spherical mechanics. The result of this section appeared already
in \cite{hkln}, but the current formulation is given in terms of $so(3)$
representations.

For any function $f$ on phase space, define the associated Hamiltonian
vector field by the Poisson bracket action $\hat f =\{f,.\}$. For example, the Hamiltonian vector fields
 corresponding to the generators $H, D, K$ (\ref{so2}), and  Casimir element \eqref{casimir} read
\begin{equation}
\label{so12-vec}
\hat H=
p_r\frac{\partial}{\partial r}
+\frac{{\ical}}{r^3}\frac{\partial}{\partial p_r}
+\frac{\hat{\ical}}{2r^2},
\qquad
\hat K=
-r\frac{\partial}{\partial p_r},
\qquad
\hat D=
r\frac{\partial}{\partial r}
-
p_r\frac{\partial}{\partial p_r},
\end{equation}
\begin{equation}
\label{hatI}
\und
\hat\ical =4{H \hat K  }+4{ K \hat H  }-2{ D \hat D}.
\end{equation}
Since the assignment $f\mapsto \hat f$ is a Lie algebra homomorphism, the vector fields $\hat H, \hat K, \hat D $
satisfy the $so(1,2)$ algebra \eqref{so120},  and  the vector field of the Casimir element $\hat\ical$,
of course, commutes with them.

Any constant of motion is the lowest weight vector of the conformal
algebra (\ref{so120}), since it is annihilated by the Hamiltonian.
Without any restriction, one can choose  it to have a certain conformal dimension (spin):
\begin{equation}
\label{hw}
\hat H I_s= 0,
\qquad
\hat D I_s= -2s I_s.
\end{equation}
A conformal mechanics which describes identical particles and possesses
a permutation-invariant cubic (in momenta, $s{=}3/2$) constant of motion
commuting with the total momentum?yields the rational Calogero model,
which is an integrable system \cite{braden}.

In the following, we consider only nonnegative integer and half-integer values of the spin $s$, so that
$I_s$ yields a finite-dimensional (nonunitary) representation
of the $so(1,2)$ algebra \eqref{so12-vec}. This includes the $N$-particle rational Calogero
model and its extensions, whose  Liouville constants of motion  are polynomials in the momenta.

Our goal is to derive the constants of motion for the
``spherical'' Hamiltonian \eqref{casimir} from the constants of motion of the
initial conformal Hamiltonian.
Using \eqref{so2}, \eqref{so12-vec}, and \eqref{hatI}
it is easy to see that the conservation condition \eqref{hw} is equivalent to the equation
\begin{equation}
\label{hatI1}
(\hat \ical-\hat M)\,I_{s}(p_r,r,u)=0,
\qquad
\text{where}
\quad
 \hat M=2(\hat S_--\ical\hat S_+).
\end{equation}
Here, the one-dimensional vector fields $\hat S_\pm$ together with $\hat S_z$
are given by
\begin{equation}
\hat S_+=\frac{1}{r}\frac{\partial}{\partial p_r},
\qquad
\hat S_-=-p_rr^2\frac{\partial}{\partial r},
\qquad
\hat S_z=-\frac{1}{2}\left(r\frac{\partial}{\partial r}
+p_r\frac{\partial}{\partial p_r}\right).
\end{equation}
Interestingly, they form an $so(3)$ algebra,
\begin{equation}
\label{so3}
[\hat S_+,\hat S_-]=2\hat S_z,
\qquad
[\hat S_z,\hat S_\pm]=\pm { \hat  S}_\pm.
\end{equation}
Note that $\hat S_+$ is generated by the Hamiltonian $ S_+=-\log(r)$ while the
other two vector fields are not Hamiltonian.

The integral \eqref{hw} can be presented as a sum of terms with decoupled
radial and angular coordinates and momenta~\footnote{
In comparison to the definition of $f_{s,m}(u)$ in \cite{hkln},
we have multiplied the binomial factor and applied an index shift
$m\to m-s$. This makes more apparent the $so(3)$ properties and
simplifies further relations.},
\be
\label{decomp}
 I_s(p_r, r, u) =\sum_{m=-s}^s f_{s,m}(u)\ R_{s,m}(p_r,r)
\qquad\textrm{with}\quad
R_{s,m}(p_r,r)=  \sqrt{\binom{2s}{s{+}m}}\,\frac{p_r^{s-m}}{r^{s+m}}.
\ee
The radial functions $R_{s,m}$  form a spin $s$-representation
($s=0,\frac12,\dots$) of the $so(3)$ algebra \eqref{so3},
\begin{equation}
\label{spin-s}
\hat S_+R_{s,m}=\sqrt{(s{-}m)(s{+}m{+}1)}\,
R_{s,m+1},
\qquad
\hat S_-R_{s,m}=\sqrt{(s{-}m{+}1)(s{+}m)}\,
R_{s,m-1},
\qquad
\hat S_zR_{s,m}=m\,R_{s,m}.
\end{equation}
Hence, $\mathcal{\hat I}$ acts nontrivially
only on the angular functions,  while the $\hat S_a$
act on the radial ones. Due  to the convolution \eqref{decomp},
one can shift the latter action to the angular functions by transposing the $so(3)$
matrices.
As a result, the action of ${\cal \hat I}$ on the spin-$s$ states
$f_{s,m}$ is given by
\be
\label{recurrent}
\hat\ical f_{s,m} = \sum_{m'}M_{mm'}f_{s,m'}
=2\bigl(\sqrt{(s{-}m)(s{+}m{+}1)}
f_{s,m+1}-{\ical}\sqrt{(s{-}m{+}1)(s{+}m)}f_{s,m-1}\bigr).
\ee
This is a system of $2s{+}1$ first-order linear homogeneous differential equations for the angular functions
$f_{s,m}(u)$. The coefficients depend only on $\ical$, which commutes with the differential operator,
and so they can be treated as constants.
Note that all angular coefficients must obey the related $(2s{+}1)$th-order linear homogeneous differential equation
\be
\label{diff-eq}
\text{Det}(\hat\ical -M)f_{s,m}=0,
\ee
which is, in fact, equivalent to the system \eqref{recurrent}, since any solution $f$ of \eqref{diff-eq}
also generates a solution of the original system.
Indeed, using \eqref{recurrent}, one can recursively express
each function $f_{s,m}$ as a $(s{\pm}m)$th-order polynomial in $\hat\ical$
acting on the function $f_{s,{\mp}s}$.
Diagonalization of the matrix  $M$ decouples the system \eqref{recurrent}
into independent equations, pertaining to the eigenvalues and eigenvectors
of the vector field $\hat\ical$.

Consider now some consequences of the relation \eqref{recurrent}.
From a constant of motion of the Hamiltonian, one can construct
other constants with the same conformal spin
by successive application of the vector field generated by the spherical Hamiltonian:
\be
I_s\stackrel{\hat\ical}{\longrightarrow} I_s^{(1)}\stackrel{\hat\ical}{\longrightarrow} I_s^{(2)}
\stackrel{\hat\ical}{\longrightarrow}\ldots\stackrel{\hat\ical}{\longrightarrow}  I_s^{(k)}
\stackrel{\hat\ical}{\longrightarrow} \ldots,
\qquad \quad I_s^{(k)}:=\hat\ical^k I_s.
\ee
In general, the members of this sequence are not in involution.
At most the first $2s{+}1$ integrals can be independent, while the
remaining ones are expressed through them linearly with $\ical$-dependent
coefficients, since the vector field $\hat\ical$ acts on the $(2s{+}1)$-vector
of constants $I_s^{(k)}$ as a square matrix with $\ical$-valued entries.
The exact amount of
functionally independent  integrals depends on the $I_s$ as well as on
the concrete realization of the conformal mechanics.

\section{Constants of motion of the spherical mechanics}
In this Section we present the construction of the constants of motion for
the spherical mechanics $(\omega_0, \ical)$ from those for the initial
conformal mechanics, based on $so(3)$ group representations.
This method yields constants of motion of any conformal dimension
and recovers the expressions found in \cite{hkln}.

Any constant of motion $I_s$ of the original Hamiltonian is given by its coefficients in the
decomposition \eqref{decomp}.
The related conservation condition \eqref{hatI1},
\eqref{recurrent}, or \eqref{diff-eq} is decoupled
into independent equations upon diagonalization of the matrix $M$,
\be
\label{hatI3}
\hat M=4\sqrt{-\ical}\; \hat U\,\hat S_z\,\hat U^{-1},
\qquad\textrm{where}\quad
\hat U=(-\ical)^{\frac12 \hat S_z}e^{-\frac{i\pi}{2} \hat S_y}
\quad\textrm{with}\quad \hat S_y=\sfrac1{2i}(\hat S_+-\hat S_-).
\ee
Thus, up to an $\ical$-valued factor, the vector field $\hat M$ is equivalent to the usual spin-$z$ projection operator.
The operator $\exp(-\frac{i\pi}2\hat S_y)$ maps $\hat S_z$ to $\hat S_x$. The latter is then transformed
to $\hat M$ by the operator $(-I)^{\hat S_z/2}$, which, for the present, means a formal power series.
Together with the factor $i\sqrt{\ical}$ it contains square roots of $\ical$. Thus $\hat M$  is, in general, complex and multi-valued.
When the potential is positive, as is the case in Calogero models, the spherical
part is strictly positive, and the operator \eqref{hatI3}
is complex but single-valued.  In any case, all square roots will cancel in
the final expressions for the constants of motion.

Define now the rotated basis for the algebra \eqref{so3}, which is formed by the eigenstates of
the operator $\hat M$. Using \eqref{hatI3}, we obtain
\be
\label{Rx}
\begin{split}
& {\widetilde R}_{s,m}=(\hat U R)_{s,m}=\sum_{m'} U_{m'm} R_{s,m'},
 \qquad
 U_{m'm} = d^s_{m'm}(\pi/2)(-\ical)^\frac{m'}2,
\\
& \hat M {\widetilde R}_{s,m}= m {\widetilde R}_{s,m},
 \end{split}
\ee
where  $d^s_{m'm}(\beta)$ is the  Wigner's small $d$-matrix, which describes
the rotation around the $y$ axis in the usual spin-$s$ representation \eqref{spin-s}.
The explicit expressions are given in the Appendix,
see \eqref{d-def}, \eqref{wigner}, \eqref{d-pi2}. There we have also
collected some  formulae and relations among the  $d$-matrix elements
which are relevant for this article.
Note that the functions ${\widetilde R}_{s,m}$ now depend on the
angular variables also through $\ical$.
The integral \eqref{decomp} of the original Hamiltonian can be presented
in terms of the rotated functions as
\be
\label{decomp2}
 I_s(p_r, r, u) =\sum_{m=-s}^s {\widetilde f}_{s,m}(u){\widetilde R}_{s,m}(p_r,r,\ical(u)).
\ee
The new coefficients are expressed in terms of old ones by means of the
inverse transformation [compare \eqref{decomp} with \eqref{decomp2} and \eqref{Rx}]:
\be
\label{fx}
 {\widetilde f}_{s,m}=\sum_{m'}U^{-1}_{mm'}f_{s,m'}
=\sum_{m'} (-\ical)^{-\frac {m'}2} d_{m'm}^s(\pi/2)f_{s,m'}.
\ee
In the second equation, we have applied the orthogonality condition of the $d$-matrix
[the first equation in \eqref{d-rel}].
Substituting the decomposition \eqref{decomp2} into \eqref{hatI1}
and using the eigenvalue-eigenfunction equation form \eqref{decomp2},
we arrive at a similar eigensystem equation for the vector field $\hat\ical$
and the rotated angular coefficients:
\be
\label{eigen}
\hat\ical {\widetilde f}_{s,m}(u)=4m\sqrt{-\ical(u)} {\widetilde f}_{s,m}(u).
\ee
This provides a formal solution to the system \eqref{recurrent}. For systems
with positive spherical part, the eigenvalue is a well-defined purely imaginary
function, and the evolution of the coefficients driven by the spherical Hamiltonian
oscillate with a frequency proportional to $m$,
\be
{\widetilde f}_{s,m}(t)=e^{iw_m(t-t_0)  }{\widetilde f}_{s,m}(t_0)
\qquad\textrm{with}\quad
\omega_m=4m\sqrt{\ical}.
\label{aa}\ee

Various combinations of these quantities give rise to constants of motion
for the spherical Hamiltonian.  In particular, for integer spin $s$,
the zero-frequency coefficient ${\widetilde f}_{s,0}(u)$ is an integral itself.
Using the explicit expression of the Wigner $d$-matrix
for this case \eqref{d-m0}, one can express it in terms of the original
coefficients:
\be
\label{int-lin}
\begin{split}
\mathcal{J}_s(u)= \ical(u)^{\frac{s}2}{\widetilde f}_{s,0}(u)=
 \sum_{m=-s}^s\frac{(s{+}m{-}1)!!(s{-}m{-}1)!!}{\sqrt{(2s)!}}\delta_{s{-}m,2\Z}\ical(u)^{\frac{s{-}m}{2}}f_{s,m}(u)
\\
=\sum_{\ell=0}^{s}\frac{(2\ell{-}1)!!(2s{-}2\ell{-}1)!!}{\sqrt{(2s)!}}\ical(u)^{\ell} f_{s,2\ell{-}s}(u).
\end{split}
\ee
Here, $\Z$ denotes the set of integer numbers, so that
$\delta_{k,2\Z}=1$ for even values of $k$ and vanishes for the odd values.
The supplementary $\ical$-dependent factor in front of the angular coefficient
eliminates the fractional powers of $\ical$, leaving only integral powers of
$\ical$ in the final expression.
Up to a normalization factor, \eqref{int-lin} coincides with the expression (5.2) of \cite{hkln}.

For integer values of~$s$,
the same integral can also be obtained from the equivalent higher-order differential
equation \eqref{diff-eq}. Indeed, due to \eqref{hatI3} or \eqref{eigen}, the related
differential operator takes the following form:
\be
\text{Det}(\hat\ical -M)=
\prod_{m=-s}^s({\hat{\ical}} -4 m \sqrt{-\ical})
=:
{\displaystyle
\begin{cases}
 \hat\ical\hat\Delta_s &
\textrm{for $s\in\Z$},
\\[5pt]
\hat\Delta_s &
\textrm{for $s\in\Z{+}\sfrac12$},
\end{cases}
}
\qquad\textrm{with}\quad
\hat\Delta_s=\prod_{0<m\le s}(\hat\ical^2 +16m^2\ical).
\label{det}\ee
Therefore, for integer spin value, \eqref{diff-eq}  is reduced to
\be
\hat\ical \hat\Delta_s f_{s,m}\equiv\prod_{m=1}^s(M^2 +16m^2\ical)f_{s,m}=0,
\ee
which implies that $\hat\Delta_s f_{s,m}$ is an integral of motion of $\ical$. The operator
$\hat\Delta_s$ projects out all but one of the eigenfunctions ${\widetilde f}_{s,m}$,
\be
\hat\Delta_s {\widetilde f}_{s,m}=\delta_{m0}(s!)^2(16\ical)^s{\widetilde f}_{s,m}.
\ee
 Therefore, the above integral has to be proportional to \eqref{int-lin}.
This can be verified independently if we apply $\hat\Delta_s$ to both
sides of the inversion of \eqref{fx} and use \eqref{Rx}, \eqref{int-lin}, \eqref{d-m0}:
\be
\label{f-fx}
\hat\Delta_s f_{s,m}
=U_{m0}\hat\Delta_s {\widetilde f}_{s,0}
=\delta_{s-m,2\Z}\,c_{s,m}\,\ical^{\frac{s+m}2} {\cal J}_s
\qquad\textrm{with}\quad
c_{s,m}=(-8i)^ss!\binom{s}{\frac{s+m}2}\sqrt{(s{-}m)!(s{+}m)!}\,.
\ee

How to construct an integral of $\ical $ from an integral of $H$ with  half-integral conformal spin?
The corresponding representation has no $m{=}0$ state, but one can consider such a state
in the integral $I_{2s}=I_s^2$, which has integral spin value equal to $2s$.
In general, integrals of
 $\ical$ can be built also by bilinear combinations of $f_{s,m}(u)$ with opposite values
 of the spin projection. In fact, any state
\be
\label{Jsm}
\begin{split}
{\cal J}_{s}^{m}
= (-\ical)^s {\widetilde f}_{s,m} {\widetilde f}_{s,-m}
&=\sum_{m',m''}
 i^{4s+m''-m'}d_{m''m}^s(\pi/2)d_{m'm}^s(\pi/2)\;
 \ical^{s-\frac{m'+m''}{2}}f_{s,m'}f_{s,m''}
 \\
&=  \sum_{m',m''}
 \delta_{m''-m',2\Z}\,(-1)^{2s+\frac{m''-m'}{2}}d_{m''m}^s(\pi/2)d_{m'm}^s(\pi/2)\;
 \ical^{s-\frac{m'+m''}{2}}f_{s,m'}f_{s,m''}
\end{split}
\ee
is  an integral of $\ical$. In the first equation, we have used the symmetry property
\eqref{d-rel-pi2} of the $d$-matrix.
The Kronecker delta appears after symmetrization over the two summation indices in the first double sum, with the help of
\be
\sfrac12(i^{m''-m'}+i^{m'-m"})
= i^{m''-m'}\,\sfrac12(1+(-1)^{m'-m''})=i^{m''-m'}\delta_{m''-m',2\Z}.
\ee
Therefore, the constant of motion ${\cal J}_{s}^{m}$ of the spherical Hamiltonian is a real polynomial of order $2s$ in $\ical$.

There is a clear interpretation of the constructed integrals in
terms of representation theory.
Take some set of angular functions satisfying
\eqref{hatI1} or \eqref{recurrent}, which means that
the related quantity $I_s$ \eqref{decomp}  is
an integral of $H$. Then, according to the tensor
product of $so(3)$ representations, one can
construct other sets of angular functions,
\be
\label{fs'm}
f_{S,m}(u)= \sum_{m_1+m_2=m} C_{s,m_1,s,m_2}^{S,m} f_{s,m_1}(u)f_{s,m_2}(u)
\qquad\textrm{with}\quad
S=2s,2s-2,\ldots,
\ee
which satisfy a similar equation.
The multiplets with odd values of $S{-}2s$ are absent in the symmetric tensor product,
due to the  exchange symmetry of the Clebsch-Gordan coefficients
\eqref{CG-exchange}.
{}From the angular functions \eqref{fs'm} one can compose ``new'' integrals of the original Hamiltonian
via
\be
\label{I'}
I'_{S}=\sum_m f_{S,m}R_{S,m}
\qquad\textrm{with}\quad
S=2s,2s-2,\ldots,
\ee
each corresponding to a symmetric multiplet in the tensor product
of two spin-$s$ multiplets.
Note that the first integral from this set
just coincides with the square of the original integral,
$I'_{2s}=I_s^2$, as can easily be verified using \eqref{CG-1}.
Since $S$ is always integer, the related multiplet contains an
$m=0$ state, which is a constant of motion of the spherical Hamiltonian:
\be
\label{Fss}
\mathcal{F}^{S}_{s}(u)
=\sum_m C_{s,m,s,-m}^{S,0}\mathcal{J}^s_m(u).
\ee
Unwanted fractional powers of $\ical$ cancel as before.
These two sets of integrals are, of course, equivalent.

A similar ``blending'' procedure can be applied to the mixing of two different integrals
$I_{s_1}$ and $I_{s_2}$ with integer value of $s_1{-}s_2$. The resulting integrals of $\ical$ are
parameterized by the whole set of $2s_{\min}{+}1$ different angular momenta obeying the sum rule.

The construction straightforwardly generalizes also to
multilinear forms composed from the angular functions.
The expression \eqref{Jsm} expands to
\be
{\cal J}^{m_1\dots m_k}_{s_1\dots s_k}(u)
=\ical(u)^{\frac12\sum_\ell s_\ell}\prod_{\ell=1}^k
{\widetilde f}_{s_\ell,m_\ell}(u)
\qquad
\textrm{with}
\quad
{\sum_{\ell=1}^km_\ell=0},
\ee
where the last relation implies that the total spin $\sum s_\ell$ must be an integer.
These observables can be combined into a single multiplet of integer spin $S$
by a $(k{-}1)$-fold application of the Clebsch-Gordan decomposition. The final set
of observables ${\widetilde f}_{S,m}$ forms an integral of the original Hamiltonian, while
its $m{=}0$ element corresponds to an integral of the spherical Hamiltonian.

So far, we have only considered products of the angular functions.
More generally however, one could also employ
fractions of them, with the same spin  projection of the numerator and
the denominator, such as ${\widetilde f}_{s_1,m}/{\widetilde f}_{s_2,m}$.
Of course, this entails introducing singularities, which might create problems
for the quantization due to inverse powers of moments.

It has to be mentioned that the variety of angular constants of motion constructed here
are not independent. It may even happen that
some of them vanish. Moreover, the compatibility of the integrals of
motion for $H$ does not at all yet imply the compatibility of the associated integrals
for $\ical$, as can be seen from \eqref{Jsm}.

\subsection*{Examples}

At the end of this section, we demonstrate our method by presenting
some simplest examples for the obtained constants of motion.

First we note that there exist two bilinear conserved quantities
\eqref{Jsm} and \eqref{Fss}, which have a rather simple form in
terms of the original angular coefficients.
The first one is the canonical ``singlet'', which is the same
both in the original and the rotated basis,
\be
\label{singlet}
\mathcal{F}^{0}_{s}(u)\sim
\sum_{m}(-1)^{s-m} {\widetilde f}_{s,m}{\widetilde f}_{s,-m}
 = \sum_{m}(-1)^{s-m} f_{s,m}f_{s,-m}.
\ee
The second one is given by the trivial superpositions of the states
\eqref{Jsm}, which is reduced by the orthogonality of the $d$-matrices
to
\be
\label{square}
\sum_m {\cal J}_s^m
\sim
\sum_{m}\ical^{s-m}f_{s,m}^2.
\ee

For the integral $I_s$ of the Hamiltonian $H$ with conformal spin $s{=}\frac12$, the
general formula \eqref{Jsm} takes its simplest form, up to a normalization factor,
\begin{equation}
\mathcal{J}_{\frac12}^{\frac12}
\sim
\ical f_{\frac12,-\frac12 }^2 +f_{\frac12,\frac12 }^2.
\end{equation}

Consider now the integral with conformal spin $s{=}1$ of the original Hamiltonian.
The related linear integral of $\ical$ is (see \eqref{int-lin})
\be
{\cal J}_{1}\sim\ical f_{1,1}+ f_{1,-1}.
\ee
In addition, there are two quadratic integrals given by \eqref{Jsm},
one of which (${\cal J}_{s=1}^{m=0}$) is the square of the above integral,
while the other one can be identified with either \eqref{singlet} or \eqref{square}.
The Hamiltonian itself can be considered as a particular case. For
$I_1=H$, the coefficient $f_{10}$ vanishes while the others become constants,
so the sole constant of $\ical$ extracted from $H$ is $\ical$ itself.

The first nontrivial case corresponds to the next conformal spin $s=\frac32$,
when there is no linear but two independent quadratic integrals.
The simplest choice then are the two functions  \eqref{singlet} and \eqref{square}.

\section{Four-particle Calogero model}

Let us use the general method developed in the previous section
to construct the complete set of constants of motion
for the spherical mechanics of the four-particle Calogero model after
decoupling the center of mass (i.e.\ of the rational $A_3$ Calogero model).
This spherical mechanics also describes a particle on the two-dimensional
sphere, interacting by the Higgs-oscillator law with force centers located
in the vertices of a cuboctahedron. By this reason, the
system was termed ``cuboctahedric Higgs oscillator''
\cite{cuboct}.

We remind that the standard rational Calogero model,
\be
\label{Calogero}
H=\frac{1}{2}\sum_{i=1}^N p_i^2 + \sum_{i<j}\frac{g^2}{(x_i-x_j)^2},
\ee
has $N$ Liouville constants of motion, given in terms of a Lax matrix by the expression \cite{polychronakos}
\be
\label{Liouville}
I_s=\text{Tr}\,L^{2s} \qquad\textrm{with}\quad s=\sfrac12,1,\dots,\sfrac{N}2,
\ee
where
\be
\label{Lax}
L_{jk}=\delta_{jk}p_k+(1{-}\delta_{jk})\frac{ig}{x_j-x_k}.
\ee
Hence,  $I_\frac12=\sum_i p_i$ and $I_1=H$.
Furthermore, the quantities
\be
\label{woj}
I_s^{(1)}=\hat\ical I_s \qquad\textrm{for}\quad \qquad s\ne 1
\ee
 coincide with Wojciechowski's additional
integrals \cite{hkln}. Together with \eqref{Liouville}, they form a complete set of
functionally independent integrals making the system maximally superintegrable \cite{woj83}.

We choose $N{=}4$ and pass to new coordinates
\be
y_0=\sfrac12(x_1{+}x_2{+}x_3{+}x_4),\qquad
y_1=\sfrac12(x_1{+}x_2{-}x_3{-}x_4),\qquad
y_2=\sfrac12(x_1{-}x_2{+}x_3{-}x_4),\qquad
y_3=\sfrac12(x_1{-}x_2{-}x_3{+}x_4)
\ee
and associated momenta $\tilde p_i$ with $i=0,1,2,3$. This transformation
decouples the center-of-mass coordinate $y_0$ and momentum $\tilde p_0$ from
the others. After setting
\be
y_0=\tilde p_0=0,
\ee
 the Hamiltonian takes the form of the rational $D_3{\sim}A_3$ Calogero model \cite{cuboct}
\be
\label{D3}
H=
\sfrac{1}{2}\sum_{i=1}^3\tilde p_i^2+\sum_{i,j=1}^3\left(\frac{g^2}{(y_i-y_j)^2}+\frac{g^2}{(y_i+y_j)^2}\right)
=  \sfrac{1}{2}p_r^2+\frac{{\ical}(p_\theta , p_\varphi, \theta
,\varphi )}{2r^2}.
\ee
In the second equation,
we introduced spherical coordinates $(r,\theta,\varphi)$ on $\R^3(y_1,y_2,y_3)$
together with their conjugate momenta $(p_r,p_\theta,p_\varphi)$, so that
\be
\ical(p_\theta,p_\varphi, \theta
,\varphi) ={p^2_\theta}+\frac{p^2_\varphi}{\sin^2\theta}+
\frac{2g^2}{\sin^2\theta}\sum_{\pm} \left[\frac{1}{(\cos \varphi
\pm\sin\varphi)^2}+\frac{1}{(\cot\theta\pm\sin\varphi )^2}
+\frac{1}{(\cot\theta\pm\cos\varphi )^2} \right], \label{z4}
\ee
in accord with the spherical symplectic structure
$\omega_0=dp_\theta \wedge d\theta  +d p_\varphi\wedge d\varphi $.

According to \eqref{Liouville} and \eqref{Lax},
the conformal Hamiltonian \eqref{D3} has two Liouville constants of motion
of conformal dimension three and four, given by
\begin{align}
\label{trL3}
\begin{split}
I_\frac32=\text{Tr}(L^3)= \sum_{i=1}^4 p_i^3 + \ldots
\quad
&= 3 \tilde p_1 \tilde p_2 \tilde p_3 + \ldots
\quad
=\sfrac{3}{2} p_r^3 \cos\theta\sin^2\theta\sin 2\varphi + \ldots\;,
\end{split}
\\
\label{trL4}
\begin{split}
I_2=\text{Tr}(L^4)=\sum_{i=1}^4 p_i^4 +  \ldots
\quad
&=\sfrac{1}{4}(\tilde p_1^4+\tilde p_2^4+\tilde p_3^4)+
\sfrac{3}{2}(\tilde p_1^2 \tilde p_1^2 + \tilde p_1^2 \tilde p_3^2 + \tilde p_2^2\tilde p_3^2 ) + \ldots
\\
&=\sfrac{1}{4} p_r^4 \left(\sin^2 2\theta +  \sin^4\theta ~ \sin^2 2\varphi + 1 \right)+\ldots\;.
\end{split}
\end{align}
Here, we have written out only the terms of highest order in the 
radial momentum.
Comparing  \eqref{trL3} and \eqref{trL4} with \eqref{decomp},
we obtain the spherical functions $f_{s,-s}$ as the coefficients 
of the monomials $p_r^{2s}$,
\be
f_{\frac{3}{2}}(\theta,\varphi)=\sfrac{3}{2}\cos\theta\sin^2\theta\sin 2\varphi,
\qquad
f_{2}(\theta,\varphi)=\sfrac{1}{4}\left(\text{sin}^2 2\theta +  \text{sin}^4\theta ~ \text{sin}^22\varphi\right).
\label{f34}
\ee
Here and in the following, we use for convenience the shorter notation
\be
f_s(\theta,\varphi):=f_{s,-s}(\theta,\varphi).
\ee
The Liouville integrals \eqref{trL3} and \eqref{trL4} are supplemented by the 
two related Wojciechowski integrals $I_\frac32^{(1)}$ and $I_2^{(1)}$ 
\eqref{woj}, whose leading-term coefficients are (see \eqref{woj})
\be
g_{\frac{3}{2}}=\hat{\ical}\,f_{\frac{3}{2}} \und
g_{2}=\hat{\ical}\,f_{2}.
\ee
Note that the $f_s$ depend on the angles only while the $g_s$ are linear
in the angular momenta.
Together with the Hamiltonian \eqref{D3}, we obtain a complete set 
$\{H,I_\frac32,I_\frac32^{(1)},I_2,I_2^{(1)}\}$ of five independent integrals.

In order to derive the Poisson algebra of integrals, we compute first 
the commutators between the related coefficients:
\be
\begin{aligned}
& \{f_{\frac{3}{2}},g_{\frac{3}{2}}\}= 18\,(f_{\frac{3}{2}}^2-f_2), \quad
&& \{f_2, g_2\} = 8\,(4f_2^2-\sfrac13{f_{\frac{3}{2}}^2}-f_2), \quad
&& \{f_{\frac{3}{2}},g_2\}= \{f_2,g_{\frac{3}{2}}\}
 = 8f_{\frac{3}{2}} (3  f_2 - 1),
\\[4pt]
& \{f_{\frac{3}{2}},f_2\}= 0,
&& \{g_{\frac{3}{2}}, g_2\} =
 4\,(2g_{\frac{3}{2}} f_2 - 3 f_{\frac{3}{2}} g_2).
&&
\end{aligned}
\ee
Since the map $I_s\to f_s$ is a Poisson algebra homomorphism \cite{hkln},  
we immediately get the analogous relations for the constants of motion by inserting
powers of $2H$ in order to balance the conformal spins on both sides of the equations 
(the coefficient for the Hamiltonian \eqref{D3} is a constant: $f_1=\sfrac12$).
Thus, the nontrivial Poisson brackets are
\be
\begin{aligned}
 & \{I_\frac32,I_\frac32^{(1)}\}= 18\,(I_\frac32^2-2 I_2 H),
  && \{I_2, I_2^{(1)}\} = 8\,(4I_2^2-\sfrac{2}{3} I_\frac32^2 H-4 I_2 H^2),
\\[4pt]
  &\{I_\frac32,I_2^{(1)}\}= \{I_2,I_\frac32^{(1)}\}= 8 I_\frac32(3 I_2- 4 H^2),
\quad
 && \{I_\frac32^{(1)}, I_2^{(1)}\} =4\,(2I_\frac32^{(1)} I_2 - 3 I_\frac32 I_2^{(1)}).
\end{aligned}
\ee
This is a particular realization of part of the quadratic algebra 
related to the Hamiltonian \cite{kuznetzov} (see \cite{sasaki01} 
for rational Calogero models based on arbitrary root systems). 
It is expressed in terms of independent generators, therefore
higher orders appear on the right-hand sides.

We now derive a complete set of functionally independent constants of motion 
for the spherical mechanics of the four-particle Calogero model.
The second expression in \eqref{det} immediately yields 
the spherical constant of motion associated with
\eqref{trL4},
\be
{\cal J}_2 = -\sfrac{1}{\sqrt{6}}\bigl(\sfrac{1}{256} \hat{\mathcal{I}}^4 + \sfrac{5}{16} \mathcal{I} \hat{\mathcal{I}}^2
+ 4 \mathcal{I}^2\bigr) f_{2}.
\ee
Its explicit expression, which can be calculated using \eqref{z4} and \eqref{f34}, is highly complicated,
\be
\label{J2}
\begin{aligned}
{\cal J}_2=&\sfrac{1}{\sqrt{6}}\biggl[
\sfrac{1}{16} (3 \cos4\varphi-11)\,p_{\theta}^4-
\sfrac{3}{4}\cot\theta \sin 4\varphi\ p_{\theta}^3 p_{\varphi}-
\Bigl(\frac{11{+}9 \cos4\varphi}{8\sin^2\theta}{+}\sfrac{9}{4}\sin^2 2\varphi\Bigr)p_{\theta}^2 p_{\varphi}^2
\\ 
+&\sfrac{3}{4} \cot^3\theta \sin 4\varphi\ p_{\theta}p_{\varphi}^3+
\frac{3 \cos ^4\theta \cos 4 \varphi + 21 \sin ^4 \theta - 18 \sin ^2 \theta -11}{16\sin^4\theta}\,p_{\varphi}^4
\biggr]
\\[4pt]
+&g^2K_1(\theta,\varphi)\,p_{\theta}^2
+g^2K_2(\theta ,\varphi)\,p_{\theta }p_{\varphi}
+g^2K_3(\theta ,\varphi)\,p_{\varphi}^2+g^4K_4(\theta ,\varphi),
\end{aligned}
\ee
where the functions $K_1(\theta ,\varphi),K_2(\theta
,\varphi), K_3(\theta ,\varphi), K_4(\theta ,\varphi)$ are given in 
Appendix C.

The system of equations \eqref{recurrent} can be applied in order to
express the coefficients $f_{\frac32,m}$ in terms of the ``lowest'' one:
\be
f_{\frac32,-\frac12}=\sfrac{1}{2\sqrt{3}}\hat\ical f_{\frac32},
\qquad
f_{\frac32,\frac12}=
\bigl(\sfrac{1}{8\sqrt{3}}\hat\ical^2+\sfrac{\sqrt{3}}{2}\ical \bigr) f_{\frac32},
\qquad
f_{\frac32,\frac32}=
\bigl(\sfrac{1}{48}\hat\ical^2+\sfrac{7}{12}\ical \bigr)\hat\ical f_{\frac32}.
\ee
Then, using \eqref{z4} and \eqref{f34},
one obtains the spherical constants of motion (\ref{Jsm}) associated with \eqref{trL3},
namely ${\cal J}_{\frac{3}{2}}^{\frac{3}{2}}$ and ${\cal J}_{\frac{3}{2}}^{\frac{1}{2}}$.
Their explicit expressions are rather lengthy:
\begin{align}
\label{J12}
\begin{split}
{\cal J}_{\frac{3}{2}}^{\frac{1}{2}} = -&\sfrac{3}{32}\sin^2 2 \varphi \,p_\theta^6
- \sfrac{3}{16}\cot \theta  \sin 4 \varphi \,p_\theta^5 p_\varphi
- \sfrac{3}{128 \sin^2 \theta} \left( 6 \cos^2 \theta + ( 13 - 3 \cos 2 \theta ) \cos 4 \varphi \right) p_\theta^4 p_\varphi^2
+ \sfrac{3}{2} \cot \theta \sin 4 \varphi\,p_\theta^3 p_\varphi^3
\\
- &\sfrac{3}{128 \sin^4 \theta} \left(22 \sin^4 \theta
 - ( 43 - 53 \cos 2 \theta ) \cos 4 \varphi \cos^2 \theta + 6 \cos 2 \theta \right) \,p_\theta^2 p_\varphi^4
- \sfrac{3}{32 \sin^5 \theta }( 7 - 9 \cos 2 \theta ) \cos^3 \theta \sin 4 \varphi \; p_\theta  p_\varphi^5
\\
- &\sfrac{3 \cos^2 \theta }{128 \sin^6 \theta} \left(  (5+ 11 \cos 4 \varphi )\sin^2 \theta
+ (2-9\cos2\theta\sin^2\theta)(1-\cos 4 \varphi ) \right)  p_\varphi^6
\ +\ \textrm{terms of lower order in $p_\theta$ and $p_\varphi$},
\end{split}
\\
\label{J32}
\begin{split}
{\cal J}_\frac{3}{2}^\frac{3}{2}=-&\sfrac{9}{32}\sin^2 2 \varphi\, p_\theta^6
- \sfrac{9}{16}  \cot \theta \sin 4 \varphi\, p_\theta^5 p_\varphi
- \sfrac{9}{64}  \left( \sfrac{5 \cos 4 \varphi + 3}{\sin^2\theta }+10 \sin^2 2 \varphi \right)p_\theta^4 p_\varphi^2
\\
- &\sfrac{9}{64 \sin^4 \theta }
\left(5 \cos ^4\theta \cos 4 \varphi+ 10 \sin^2 \theta - 5 \sin^4 \theta  + 3 \right)
p_\theta^2 p_\varphi^4
+ \sfrac{9}{16}\cot^5\theta \sin 4 \varphi \, p_\theta  p_\varphi^5
\\
+ &\sfrac{9\cos^2 \theta}{64 \sin^6 \theta }
\left(\cos^4\theta \cos 4 \varphi- 6\sin^2 \theta - \sin^4 \theta  - 1 \right)
p_\varphi^6\ +\
\textrm{terms of lower order in $p_\theta$ and $p_\varphi$}.
\end{split}
\end{align}
Clearly, ${\cal I}$, ${\cal J}_2$, ${\cal J}_{\frac{3}{2}}^{\frac{1}{2}}$ and
${\cal J}_{\frac{3}{2}}^{\frac{3}{2}}$ cannot be functionally
independent, since our spherical mechanics has a four-dimensional phase space.
Indeed, using {\tt Mathematica}, we uncover the following algebraic relation,
\be
{\cal J}_{\frac{3}{2}}^{\frac{3}{2}} = \sfrac{1}{3}{\cal J}_{\frac{3}{2}}^{\frac{1}{2}}
+2\sqrt{\sfrac{2}{3}}{\cal J}_2 \mathcal{I}+\sfrac{1}{3}\mathcal{I}^3+
4 g^2 \mathcal{I}^2.
\ee
This is the only relation among the four constants of motion,
since \eqref{J12} and \eqref{J32}
are not in involution with  \eqref{J2}. Even their free-particle
parts ($g{=}0$ projects to the terms of highest order in the momenta)
do not commute as is easy to verify.
Hence, we have found three functionally independent spherical constants 
of motion for the $A_3$ Calogero model.
This confirms the superintegrability of that system.

\section{Conclusion}
In the present paper we have developed a general approach to the constants of motion for conformal mechanics, based on $so(3)$ representation theory.
In particular, we gave an explicit construction of the
(overcomplete set of) constants of motion for the spherical part of conformal
mechanics (``spherical mechanics''), which are related to the constants
of motion for the initial conformal system.
We have illustrated the effectiveness of our method on the example of the
rational $A_3$ Calogero model and its spherical mechanics
(which defines the cuboctahedric Higgs oscillator).
For the latter we have constructed a complete set of functionally independent
constants of motion, proving its intuitively obvious superintegrability.

Unfortunately, our approach does not allow one to select a commuting subset of
constants of motion for the spherical mechanics.
Also, it does not provide us with a rule for selecting {\it a priori\/}
functionally independent constants of motions.
Hopefully, further development of this approach will provide answers to
these questions.\\

{\large Acknowledgments.}
 We are  grateful to Vahagn Yeghikyan   for useful discussions
and comments. The work was supported by
 and ANSEF-2229PS grant and by
 Volkswagen Foundation  grant I/84~496.

\appendix
\section{Wigner (small) $d$-matrix}

The spin-$j$ representation of the rotation group parameterized by
three Euler angles is given by the Wigner $D$-matrix \cite{wigner,angular}.
We only need the (small) $d$-matrix, 
which describes the rotation around the $y$ axis,
\begin{equation}
\label{d-def}
d_{m'm}^s(\beta)=\langle sm'|\exp(-i\beta S_y) |sm\rangle,
\end{equation}
where $m,m'=-s,\dots,s$ are the spin $z$-projection quantum numbers.
Its elements are real and given by \cite{wigner}
\begin{equation}
d_{m'm}^s(\beta)=\sum_t (-1)^{t+m'-m} 
\frac{\sqrt{(s{+}m')!(s{-}m')!(s{+}m)!(s{-}m)!}}
{(j{+}m{-}t)!(m'{-}m{+}t)!(j{-}m'{-}t)!}
\left(\cos\sfrac{\beta}{2}\right)^{2s+m-m'-2t}
\left(\sin\sfrac{\beta}{2}\right)^{m'-m+2t},
\end{equation}
where the sum is over such values of $t$ that the factorials in the denominator
are nonnegative. The elements obey
\begin{equation}
\label{d-rel}
d_{m'm}^s(\beta)=d_{mm'}^s(-\beta)=(-1)^{m-m'}d_{mm'}^s(\beta)=d_{-m-m'}^s(\beta).
\end{equation}
For $\beta=\pi/2$, the above expression simplifies to
\begin{equation}
\label{d-pi2}
d_{m'm}^s(\pi/ 2)= 2^{-s} \sum_t (-1)^{t+m'-m}
\frac{\sqrt{(s{+}m')!(s{-}m')!(s{+}m)!(s{-}m)!}}
{(s{+}m{-}t)!(m'{-}m{+}t)!(s{-}m'{-}t)!}\;.
\end{equation}
Further simplifications occur when one of the spin-projection quantum
numbers vanishes, which is possible for integer spins only:
\begin{equation}
\label{d-m0}
\begin{split}
d_{m0}^s(\pi/2)&=(-1)^\frac{s+m}{2} \delta_{s-m,2\Z}
\frac{\sqrt{(s{-}m)!(s{+}m)!}}{2^s\left(\frac{s+m}{2}\right)!\left(\frac{s-m}{2}\right)!}
=(-1)^\frac{s+m}{2} \delta_{s-m,2\Z}
\sqrt{\frac{(s{+}m{-}1)!!(s{-}m{-}1)!!}{(s{+}m)!!(s{-}m)!!}}
\\
d_{0m}^s(\pi/2)&=(-1)^\frac{s-m}{2} \delta_{s-m,2\Z}
\frac{\sqrt{(s{-}m)!({s}+m)!}}{2^s\left(\frac{s+m}{2}\right)!\left(\frac{s-m}{2}\right)!}
=(-1)^\frac{s-m}{2} \delta_{s-m,2\Z}
\sqrt{\frac{(s{+}m{-}1)!!(s{-}m{-}1)!!}{(s{+}m)!!(s{-}m)!!}}
\\
d_{00}^s(\pi/2)&=(-1)^\frac{s}{2} \delta_{s,2\Z} \frac{(s{-}1)!!}{s!!}
\end{split}
\end{equation}
The factor $\delta_{s-m,2\Z}$ excludes odd values of $s{-}m$, for which the
matrix elements vanish.
For $\beta=\pi/2$, the relations are supplemented by
\be
\label{d-rel-pi2}
d_{m'm}^s(\pi/2)=(-1)^{s+m'}d_{m'-m}^s(\pi/2)=(-1)^{s-m}d_{-m'm}^s(\pi/2),
\ee
which can be obtained from $d^s_{m'm}(\pi)=(-1)^{s-m}\delta_{m',-m}$.

\section{Clebsch-Gordan Coefficients}

Clebsch-Gordan coefficients are the expansion coefficients of total-spin
eigenstates $|SM\rangle$ in terms of the product basis $|s_1m_1s_2m_2\rangle$
of eigenstates of the two coupled spins,
\be
\label{CG-def}
C_{s_1,m_1,s_2,m_2}^{S,M}=\langle s_1m_1s_2m_2|SM\rangle.
\ee
The general expression is complicated, but special cases are often quite simple
like for the highest total-spin value:
\be
\label{CG-1}
  C_{s_1,m_1,s_2,m_2}^{s_1+s_2,m_1+m_2}=
  \sqrt{\frac{
  \binom{2s_1}{s_1-m_1}\binom{2s_2}{s_2-m_2}}{\binom{2s_1+2s_2}{s_1+s_2-m_1-m_2}
  }}.
\ee
The Clebsch-Gordan coefficients have an even-odd exchange symmetry
depending on the total-spin value,
\be
\label{CG-exchange}
C_{s_1,m_1,s_2,m_2}^{s,m}=(-1)^{s_1+s_2-s} C_{s_2,m_2,s_1,m_1}^{s,m}.
\ee

\section{Coefficients of ${\cal J}_2$}
Below we write down the explicit expressions for the functions
$K_1,K_2,K_3,K_4$, which appear in (\ref{J2}).
\begin{align}
K_1(\theta,\varphi)& = \frac{1}{2^{14}\sqrt{6}\cos^2 2\varphi \
(\cos^2\theta -\sin^2\theta \cos^2\varphi)^2
(2\sin^2\theta \cos 2\varphi +3\cos 2\theta +1)}\ \times \\ \nonumber
\Bigl(&768\,(25 + 29 \cos 2 \theta) \sin^6\theta \cos 12 \varphi
+96\,(1370 + 2327 \cos 2 \theta +1542 \cos 4 \theta
+393 \cos 6 \theta) \sin^2\theta \cos 8 \varphi \\ \nonumber
&-(119258 + 175774 \cos 2 \theta +45096 \cos 4 \theta +57723 \cos 6 \theta
-10242 \cos 8 \theta +5607 \cos 10 \theta) \sin^{-2}\theta \cos 4 \varphi
\\ \nonumber
&+(1021064 + 365088 \cos 2\theta - 223008 \cos 4\theta
- 183840 \cos 6\theta - 61800 \cos 8\theta -655360\sin^{-2}\theta) \Bigr)\;,
\end{align}
\begin{align}
K_2(\theta,\varphi) &= \frac{3\cot \theta  \tan 2 \varphi }
{8\sqrt{6}\sin^2\theta\,(17\cos 4\theta+28\cos 2\theta-8\sin^4\theta\cos 4\varphi+19)^2}
\ \times \\ \nonumber
\Bigl(&351 \cos 10 \theta +1350 \cos 8 \theta +13779 \cos 6 \theta
+9992\cos 4\theta+35022\cos 2\theta-13824\sin^8\theta\cos^2\theta\cos 8\varphi
+5042 \\ \nonumber
&-64\,(81\cos 6\theta+702\cos 4\theta+1071\cos 2\theta+962)
\sin ^4\theta \cos 4 \varphi \Bigr)\:,
\end{align}
\begin{align}
K_3(\theta,\varphi) &= \frac{1 }{16\sqrt{6}\cos^2 2\varphi\,
(17\cos 4\theta+28\cos 2\theta-8\sin^4\theta\cos 4\varphi+19)^2}
\ \times \\ \nonumber
\Bigl(&162\,(13 \sin 2\varphi +\sin 6\varphi)^2\cos 8\theta
+24\,(3898{-}1569\cos 4\varphi{-}282\cos 8\varphi{+}\cos 12\varphi)\cos 6\theta
\\ \nonumber
&+36\,(6686{+}1931\cos 4\varphi{-}430\cos 8\varphi{+}5\cos 12\varphi)\cos 4\theta
+72\,(546{+}10587\cos 4\varphi{-}898\cos 8\varphi{+}5\cos 12\varphi)\cos 2\theta\\[6pt] \nonumber
&-(1087746{-}1625907\cos4\varphi{+}46158\cos 8\varphi{+}483\cos 12\varphi)
\\ \nonumber
&+262144\,(5{-}4 \cos 4\varphi)\,\sin^{-2}\theta
-32768\,(11{-}3 \cos 4\varphi)\,\sin^{-4}\theta \Bigr)\;,
\end{align}
\begin{align}
K_4(\theta,\varphi)&= \frac{-1}{64\sqrt{6} \left((60 \cos 2 \theta +33 \cos 4\theta +35) \cos2 \varphi -8 \sin ^4\theta
\cos 6 \varphi\right)^4}
\times \\\nonumber
&\Bigl[  64 (335698872 \cos 2 \theta
+204278376 \cos 4 \theta +100740648 \cos 6 \theta +30799596 \cos 8
\theta +3629304 \cos 10 \theta
\\\nonumber
&+515160 \cos 12 \theta -649944 \cos
14 \theta -194643 \cos 16 \theta +197597863) \cos 8 \varphi
\\\nonumber
&+384 \sin^4
\theta  \bigl(
(-16777208 \cos 2 \theta -15290507 \cos 4 \theta
-10272396 \cos 6 \theta -4824234 \cos 8 \theta -2019708 \cos 10
\theta \\\nonumber
&-312741 \cos 12 \theta -8174886) \cos 12 \varphi -768 \sin^8
\theta  (828 \cos 2 \theta +243 \cos 4 \theta +617) \cos 20 \varphi
\\\nonumber
&-32
\sin ^4 \theta  (290832 \cos 2 \theta +188916 \cos 4 \theta +81648
\cos 6 \theta +13851 \cos 8 \theta +166129) \cos 16 \varphi\bigr)
\\\nonumber
&+\sin^{-4} \theta \bigl(-(9941103400 \cos 2 \theta +11541549238
\cos 4 \theta +10411072176 \cos 6 \theta +8259070392 \cos 8 \theta\\\nonumber
&+4658511600 \cos 10 \theta +1965778311 \cos 12 \theta +569460204
\cos 14 \theta +67528026 \cos 16 \theta -29495988 \cos 18 \theta\\\nonumber
&-8028477 \cos 20 \theta +4163058670) \cos 4 \varphi +62158979032 \cos 2
\theta +46026533130 \cos 4 \theta +27521060688 \cos 6 \theta\\\nonumber
&+12943186248 \cos 8 \theta +4533912336 \cos 10 \theta +1033949913
\cos 12 \theta -11388780 \cos 14 \theta -94673178 \cos 16 \theta\\\nonumber
&-31001292 \cos 18 \theta -6738147 \cos 20 \theta +34904741074\bigr)\Bigr] \;.
\end{align}

\end{document}